\documentstyle[12pt,a4,epsf]{article}
\normalsize
\def\simless{\mathbin{\lower 3pt\hbox{$\rlap{\raise 5pt\hbox{$\char'074$}}
\mathchar"7218$}}}
\def\simgreat{\mathbin{\lower 3pt\hbox{$\rlap{\raise 5pt \hbox{$\char'076$}}
\mathchar"7218$}}}
\catcode`@=11
\def\beqra{\begin{eqnarray}} \def\eeqra{\end{eqnarray}}
\def\beq{\begin{equation}}      \def\eeq{\end{equation}}
\def\ds{\displaystyle}

\def\fo{\hbox{{1}\kern-.25em\hbox{l}}}

\def\ch{\@startsection{section}{1}{\z@}{-3ex plus-1ex minus-.2ex}%
        {2ex plus.2ex}{\large\sc}}

\def\; \lapp \;{\raisebox{-.4ex}{\rlap{$\sim$}} \raisebox{.4ex}{$<$}}

\def\con{\ifmmode \hbox{\bf*} \else{\bf*}\fi}   
\def\scon{\ifmmode \hbox{\footnotesize\rm\bf*} \else{\footnotesize\rm\bf*}\fi}

\def\0#1{\relax\ifmmode\mathaccent"7017{#1}
        \else\accent23#1\relax\fi}              

\def\eslash{\not{\hbox{\kern-2pt $E$}}}
\newcommand{\is}{\int \!\!\!\!\!\! \mbox{$\sum$}}

\catcode`@=12
\begin{document}
\hoffset=-0.4cm
\voffset=-1truecm
\normalsize


\begin{titlepage}
\begin{flushright}
DESY 95-109\\
\end{flushright}
\vspace{24pt}

\centerline{\Large {\bf Particle Currents in a Space-time dependent and }}
\vskip 12pt
\centerline{\Large{\bf  $CP$-
Violating Higgs Background: a Field Theory Approach}}
\vspace{24pt}
\begin{center}
{\large\bf D. Comelli$^{a,}$\footnote{Email:
comelli@evalvx.ific.uv.es. Work supported by Ministerio de
Educacion y Ciencia de Espa\~{n}a}
, M. Pietroni$^{b,}$\footnote{Email: pietroni@vxdesy.desy.de}
 and A. Riotto$^{c,d,}$\footnote{Email:riotto@tsmi19.sissa.it.
Address after November 95: Theoretical Astrophysics Group,
NASA/Fermilab, Batavia, Il 60510}}
\end{center}
\vskip 0.2 cm
\centerline{\it $^{(a)}$Departamento de Fisica Teorica,
Universitad de Valencia,}
\centerline{\it E-46110 Burjassot, Valencia, Spain}
\vskip 0.2 cm
\centerline{\it $^{(b)}$Deutsches Elektronen-Synchrotron DESY,}
\centerline{\it Notkestr. 85, D-22603 Hamburg, Germany.}
\vskip 0.2 cm
\centerline{\it $^{(c)}$Istituto Nazionale di Fisica Nucleare,}
\centerline{\it Sezione di Padova, 35100 Padua, Italy.}
\vskip 0.2 cm
\centerline{\it $^{(d)}$International School for Advanced Studies, SISSA-ISAS}
\centerline{\it Strada Costiera 11, I-34014 Miramare, Trieste, Italy.}
\vskip 0.5 cm
\centerline{\large\bf Abstract}
\vskip 0.2 cm
\baselineskip=15pt
Motivated by cosmological applications like electroweak baryogenesis, we
develop a field theoretic approach to the computation of particle currents on
a space-time dependent and CP-violating Higgs background. We consider the
Standard Model model with two Higgs doublets and CP violation in the scalar
sector, and compute both fermionic and Higgs currents by means of an expansion
in the background fields. We discuss the gauge dependence of the results and
the renormalization of the current operators, showing that in the limit of
local equilibrium, no extra renormalization conditions are needed in order to
specify the system completely.

\end{titlepage}

\baselineskip=18pt
\setcounter{page}{1}
\setcounter{footnote}{0}
{\large\bf 1. Introduction}
\vskip 0.5cm
In a recent letter, we have presented our results for the computation of
the expectation value of particle currents in a two Higgs doublet model, in
the case in which the Higgs field background is space time dependent and
CP-violating \cite{expansion}.
The original motivation for such a computation comes from the so called
spontaneous baryogenesis scenario \cite{CKN} at the electroweak phase
transition. The basic ingredients of this mechanism are CP violation in the
Higgs sector, which requires at least two Higgs doublets, and thick bubble
walls, so that at least some of  the more rapid processes (top Yukawa and
 gauge flavour diagonal interactions) can be assumed to be in local
equilibrium in the background of the electroweak bubble walls and a adiabatic
approximation can be applied. Of course, also in this scenario one must
require that the baryon number violating processes go quickly out of
equilibrium in the broken symmetry phase, which seems by now very hard to
achieve, at least in the Standard Model \cite{SM}
and in the minimal supersymmetric
extension of it \cite{hispano}.

Both in the case of explicit and
spontaneous \cite{scpv} $CP$-violation, a space-time dependent
relative phase $\delta(x)$ between the two Higgs fields can be present
inside the wall of the expanding bubble. Due to the fact that the background
is not space-time translational invariant,  the phase $\delta(x)$
can give rise locally to nonvanishing expectation values for the
particle currents present in the theory.
If baryon number violation
is then in equilibrium, these nonvanishing particle currents
can rearrange to generate a baryon asymmetry.

In order to reliably compute the final baryon asymmetry, it is then necessary,
as a first step, to determine the values of the currents induced by the
background. In the approach of refs. \cite{CKN, diffusion}
a rotation of the fermionic
fields is done to make
the Yukawa couplings real. As a consequence, a derivative coupling of
the form
\begin{equation}
{\cal L}_{int}\sim \partial_\mu\delta J^\mu,
\end{equation}
where $J^\mu$ is the current corresponding to the rotation, is induced
from the kinetic terms. In the original paper \cite{CKN}
only the time derivative $\dot{\delta}$ was taken into account.
 $\dot{\delta}$ acts as an effective chemical potential, called
"charge potential", and particle densities acquire nonvanishing
expectation values $n_i\sim q_i\: \dot{\delta}\:T^2$, where $q_i$ is the
charge of the $i$-th particle under the given rotation. The
thermodynamical evolution of the system in presence of baryon number
violation then leads to a nonvanishing baryon asymmetry. The gradient
part of $\delta(x)$ in eq. (1) was subsequently taken into account
in ref. \cite{diffusion} where the role of particle diffusion was also
discussed. Nevertheless, as we already pointed our in \cite{expansion},
using the
interaction term in eq.
(1) as a starting point to compute the perturbations to the thermal
averages presents some drawbacks.

Since the phase $\delta$ is communicated from the Higgs to the
fermion sector through the Yukawa interactions, any perturbation in the fermion
densities $n_i$ should vanish in the limit of zero Yukawa couplings $h_i$.
Also, they
should vanish in the limit of zero vacuum expectation value for the Higgs
fields $H_i(x)=v_i(x)\exp[i\theta_i(x)]$
because no spontaneous CP violation is
present in the Higgs sector in
this limit. Naively, one could then expect a suppression factor of order
$(h_i^2  v^2_i(x)/T^2)$, where $h_i$ is the relevant Yukawa coupling,
for the perturbations in the fermionic particle number
 with respect
to the original result. Since one is interested in regions of
the bubble wall where
sphalerons are still active, {\it i.e} for values of $v_i(x)/T$  typically
smaller than
one, then the above mentioned suppression factor might be crucial
 for the scenario of baryogenesis outlined above \cite{expansion}.

The reason why these suppressions do not appear in the original
treatment, is that considering eq. (1) as the only effect of the
background is equivalent to perturbing around the Higgs field configuration
 $\delta(x)=0$, $v_i(x)\neq 0$, which is not a solution of the field
equations. On the other hand, taking the field equations into account, one
immediately sees that the expected suppression factors are recovered also in
the original treatment, so that the result comes out to be rotation
independent, as it should, also in this context. Indeed, by partially
integrating eq. (1), we obtain a `perturbation term' which is given by the
hypercharge violating terms in the lagrangian, coming from the fermionic mass
terms and from the Higgs potential.
Also, from the field equations one can  see that
$\partial_{\mu}\delta(x)$ vanishes as $v_i(x)^2$ for vanishing $v_i(x)$.
In other words, it is the perturbation (1) itself, if properly considered, to
be vanishing in the limit of vanishing top Yukawa and Higgs couplings, or in
the limit of vanishing values for the background Higgs fields.

In the limit in which some interaction rates (like for example those for the
top Yukawa processes) are so high as to be in thermal equilibrium inside the
bubble walls, the local equilibrium values for the corresponding particle
numbers are of course independent on the reaction rates.

On the other hand, since their values are fixed by that of the perturbation
term (1), they are still suppressed by the above mentioned factor of $(h_t^2
v(x)^2/T^2)$.

In this paper we want to describe in more detail our approach to calculate
the expectation values of a composite operator
$\hat{O}(z)$ in a $CP$-violating
Higgs
space-time dependent background. For definitness, we will work on the
background of a bubble wall, as the one described above, however the
formalism is quite general and can be easily extended to consider other
interesting situations, like for example a non  trivial background also for
the gauge fields.

The method relies on a functional expansion
in  power
series of $\Phi_i^c(x)$ where the coefficients of the expansion are the
$n$-point 1PI Green's functions with one insertion of the
operator $\hat{O}(z)$ computed in the {\it unbroken} phase.

We shall
discuss the dependence of the expectation values from the gauge in which
the background $\Phi_i^c(x)$ is expressed, showing explicitely that the
expectation values of  gauge
invariant operators are gauge independent. Moreover, the
problem of current renormalization will be addressed.
In general, the computation of Green's functions with insertion of a
composite operator requires the introduction of new counterterms besides
those necessary for the renormalization of the basic lagrangian.
This fact leads to the well known phenomenon of the mixing
among the different renormalized currents of the theory. We shall
discuss the mixing matrix for the renormalized currents
present in the Standard Model model with two Higgs doublet  and
use the non renormalization properties for the conserved currents in order to
reduce the number of independent counterterms.

Finally, we will discuss the limit of local equilibrium, showing that, in this
case, the only expectation values that can be consistently computed are those
for the conserved currents, so that no new renormalization condition is
needed to specify the system in this limit.

The paper is organized as follows: in sect. {\bf 2} our field theory
method is described on general grounds and particular attention is given to
the classical equations of motions and to the question  of
gauge invariance. In sect. {\bf 3} the explicit calculations
for the fermionic and Higgs currents in the specific model under
consideration are given in details. They are relevant for the
spontaneous baryogenesis mechanism and their consequence have been
discussed in ref. \cite{expansion}.
Sect. {\bf 4} deals with the issue of current
renormalization, whereas the role of the conserved currents in the
thermodynamical limit is described in sect. {\bf 5}. Finally, sect. {\bf
6} presents our conclusions.
\vskip 1cm

{\large\bf 2. The expansion.}
\vskip 0.5cm

In this paragraph we will discuss our approach to the
computation of the expectation value
of an operator $\hat{O}(z)$ on a
non-zero background for the fields of the theory.

Our starting point is the finite temperature
 generating functional for the
1PI Green's functions
with insertion of an operator $\hat{O}(z)$
(in the following $\hat{O}(z)$ will represent a
particle current)
\beq
\Gamma\left[\Phi^c_i(x),\:\Delta(x)\right] = W\left[J_i(x),\:\Delta(x)\right] -
\sum_j \int d^4x J_j(x) \Phi^c_j(x),
\label{start}
\eeq
where $\Phi^c_i(x)$ are the  classical
 fields of the theory and $J_i(x)$ the
corresponding sources, while $\Delta(x)$ is the source for the operator
$\hat{O}(x)$.  Note that the Legendre transformation has been performed only
on the
fields  and not on the operator.

The quantity we are interested in is the expectation value of the operator
$\hat{O}(z)$ on the background given by the fields $\Phi^c_i(x)$, which we
will specify later,
\beq
\langle \hat{O}(z) \rangle_{\Phi^c_i(x)} = \left.
\frac{\delta \Gamma\left[\Phi^c_i,\:\Delta\right]}
{\delta \Delta(z)}\right|_{\Delta=0}
\equiv {\cal O}\left[\Phi^c_i(x)\right](z).
\label{funct}
\eeq
We can expand the functional ${\cal O}\left[\Phi^c_i(x)\right](z)$ in a
power series of $\Phi_i^c$
\beq
{\cal O}\left[\Phi^c_i(x)\right](z)=\sum_{n=0}^{\infty}
\sum_{i_1,\ldots,i_n} \frac{1}{n!} \int d^4 x_1\cdots d^4 x_n
{\cal O}_{i_1,\ldots,i_n}^{(n)}(x_1,\cdots,x_n;\:z)\Phi_{i_1}^c(x_1)\cdots
\Phi_{i_n}^c(x_n),
\label{exp}
\eeq
where the coefficients of the expansion are the n-point 1PI Green's
functions with one insertion of the operator $\hat{O}(z)$ computed in the
{\it unbroken} phase
\beq
{\cal
O}_{i_1,\ldots,i_n}^{(n)}(x_1,\cdots,x_n;\:z)=
\frac{1}{i}\left.\frac{\delta^{n+1}
\Gamma\left[\Phi^c_i,\:\Delta\right]}{\delta \Phi^c_{i_1}(x_1)\cdots
\delta \Phi^c_{i_n}(x_n)\delta \Delta(z)}\right|_{\Phi^c_i=\Delta=0}.
\label{1p1}
\eeq

Now we have to specify the background fields, $\Phi^c_i(x)$, which give the
starting point for the expansion in eq. (\ref{exp}). A priori, the relevant
background is given by the solution of the full field equations,
\beq
\left.\frac{\delta\Gamma\left[\Phi_i^c,\:\Delta=0\right]}{\delta
\Phi_i^c(x)}\right|_{\Phi_i^c=\bar{\Phi}_i^c}=0
\label{fe}
\eeq
with appropriate boundary conditions. In practice, however, we have to
consider some approximation of this complete, and unknown, solution. A
possibility could be to consider the solutions of the classical equations of
motion as the zero order approximation. However in many interesting cases the
radiative corrections to the effective potential, either at zero or at finite
temperature, are crucial in order to determine the shape of the potential and
consequently of the background. A typical example is that of a first order
phase transition induced by finite temperature corrections, in which the
existence of bubble solutions is due to the $T \neq 0$ radiative corrections
which induce a second minima in the effective potential. Following refs.
\cite{Weinberg, Gleiser}, we will then consider the classical equations of
motion in which the tree level  effective potential is replaced by the
radiatively corrected one.

In the following, we will consider the Standard Model with two Higgs
doublets, $H_1$ and $H_2$. $H_1$ has $U(1)_Y$ charge $-1/2$, and couples to
the `down-type' right handed fermions, whereas $H_2$ (charge $1/2$) couples to
the `up-type' ones.
Moreover, since only top and
bottom quarks will be relevant in the following we will not consider the
lighter quarks and leptons.

Neglecting all the Yukawa couplings but the one for the right handed top,
$h_t$,
the classical lagrangian is given by
\beqra
{\cal L}& =& -\frac{1}{4} F_{\mu \nu} F^{\mu \nu} +
\left({\cal D}_\mu H_1 \right)^\dagger {\cal D}^\mu H_1 +
\left({\cal D}_\mu H_2 \right)^\dagger {\cal D}^\mu H_2 \\
&&+ i \overline{t}_L \gamma_\mu{\not \!\! {\cal D}}^\mu t_L
+i \overline{t}_R \gamma_\mu{\not \!\!{\cal D}}^\mu t_R + (t \rightarrow b)
+ h_t \left(H_2^0 \overline{t}_R t_L - H^+  \overline{t}_R b_L\right)
 - V(H_1, H_2),
\label{lagr}
\eeqra
where the most general tree level scalar potential is given by
\begin{eqnarray}
V &=& {m_{1}}^{2}|H_1|^2 + {m_{2}}^{2}|H_2|^2 - ({m_{3}}^{2} H_1 H_2 +
{\rm h.c.})
+\lambda_1 |H_1|^4 +\lambda_2 |H_2|^4  \nonumber\\
&+& \lambda_3 |H_1|^2 |H_2|^2 + \lambda_4|H_1 H_2|^2
+\left[\lambda_5 (H_1 H_2)^2  + \lambda_6 |H_1|^2 H_1 H_2 +\lambda_7 |H_2|^2
H_1 H_2 + {\rm h.c.}\right].
\label{pot}
\end{eqnarray}
In eq. (\ref{lagr}) we have also omitted tha SU(2) gauge part, which will not
be relevant in the following.
Note that of the two phases of the Higgs fields, $\theta_1$ and $\theta_2$,
defined by $H_i(x) = v_i(x) \exp [1 \theta_i(x)]$,
only the gauge invariant combination
$\delta=\theta_1 + \theta_2$ appears in the scalar potential. The orthogonal
combination, $\chi=\theta_1 - \theta_2$, is the gauge phase, corresponding
to the would be Goldstone boson after spontaneous symmetry breaking.
One can now write down the equations of motion coming from the lagrangian in
(\ref{lagr}), with the tree level scalar potential $V(H_1, H_2)$ replaced by
the (finite temperature) radiatively corrected one, $\overline{V}(H_1, H_2)$.
We assume that the parameters of the effective potential (and the temperature)
are such that it has only one minimum in the charged directions, given by
$H^+=H^-=0$, whereas there are two minima in the neutral Higgs directions.
Then, a bubble solution exists, with $v_{1,2}(x)$ changing from a non-zero
value $v_{1,2}^+$ inside to zero outside, and $H^\pm=0$ everywhere.

At zero order, we will also take the fermion fields and currents to be zero, so
that the background is given by the solution of the following field equations:
\beqra
&&\partial^2 v_{1,2} + \frac{\partial \overline{V}}{\partial v_{1,2}}
-\frac{1}{4} v_{1,2}
\left(\partial_\mu \delta \mp {\cal D}_\mu \chi\right)^2 = 0\\
&&\partial_\mu F^{\mu \nu} = \frac{e}{4}\left[(v_1^2 - v_2^2)\partial^\nu
\delta
- (v_1^2 + v_2^2){\cal D}^\nu \chi\right] \label{gauss}\\
&&\frac{1}{4} \partial_{\nu}\left[(v_1^2 + v_2^2)\partial^\nu \delta
- (v_1^2 - v_2^2){\cal D}^\nu \chi\right] =
-\frac{\partial \overline{V}}{\partial \delta}\label{phase},
\eeqra
where the gauge invariant quantity ${\cal D}_\mu \chi$ is given by
\beq {\cal D}_\mu \chi=\partial_\mu\chi + e A_\mu.
\eeq

Assuming also that there are no electric or magnetic fields at zero order,
{\it i.e.} $F_{\mu \nu}=0$, and chosing the unitary gauge $A_\mu =0$
(see below),
we obtain from
eq. (\ref{gauss})
\beq
v_1^2 \partial_\mu \theta_1 = v_2^2 \partial_\mu
\theta_2,\:\:\:\:\:\:\:(A_\mu=0)
\label{aa}
\eeq
so that eq. (\ref{phase}) now reads
\beq
\partial^\mu\left( v_1^2 \partial_\mu \theta_1\right) =
-\frac{\partial \overline{V}}{\partial \delta}.
\label{bb}
\eeq

Inserting the solutions of the above equations into the expansion in eq.
(\ref{exp}) we will now be able to compute the various contributions to the
expectation values of the currents as a loop expansion. In the
following paragraphs  we will see that the
Higgs currents get a contribution already at the tree level, whereas,
consinstently with our assumptions, the contributions to the fermionic
currents, and to $\langle \partial_\mu F^{\mu \nu}\rangle$, arise only at one
loop.

Before concluding this section, we want to show explicitly that the
expectation value for a gauge invariant operator,
 as obtained by the expansion in
eq. (\ref{exp}), is independent on the gauge in which the classical
background has been computed. In our discussion above, we have looked for a
solution of the field equations with no electromagnetic fields, {\it i.e.}
with $F_{\mu \nu}=0$, so that it was possible to chose a gauge in which
$A_\mu=0$ everywhere (see eqs. (\ref{aa},\ref{bb})). Chosing
 a different gauge gives
rise to a different background, in which $A_\mu$ is diferent from zero, and we
want to show that the result for the gauge invariant operator $\hat{O}(z)$ is
the same as in the original background.

Following the usual procedure for the derivation of the Ward identities it is
straightforward to obtain the relation
\beq
\frac{1}{\alpha} \partial^2 \partial_\mu a^{\mu} +\partial_\mu
\frac{\delta \Gamma}{\delta a_\mu}
 - i \frac{e}{2}\left(H_1^0 \frac{\delta \Gamma}{\delta H_1^0} -
{H_1^0}^* \frac{\delta \Gamma}{\delta {H_1^0}^*}
-H_2^0 \frac{\delta \Gamma}{\delta H_2^0}
+ {H_2^0}^* \frac{\delta \Gamma}{\delta {H_2^0}^*}\right) = 0
\eeq
where $\alpha$ is the gauge parameter,
$a_\mu$, $H_{1,2}^0$, and ${H_{1,2}^0}^*$ are the classical gauge and
neutral Higgs fields, and we have put the charged Higgses and the fermion
classical fields to zero.
Differentiating the above expression with respect to the source $\Delta(x)$
keeping the background fields fixed,
and then putting $\Delta(x)=0$ we obtain the relevant identity containing the
functional ${\cal O}$ of eq. (\ref{funct}),
\beq
\partial_\mu \frac{\delta {\cal O}}{\delta a_\mu} -i\frac{e}{2} \left(
H_1^0 \frac{\delta {\cal O}}{\delta H_1^0} -
{H_1^0}^* \frac{\delta {\cal O}}{\delta {H_1^0}^*} -
H_2^0 \frac{\delta {\cal O}}{\delta H_2^0}+
{H_2^0}^* \frac{\delta {\cal O}}{\delta {H_2^0}^*}\right) = 0.
\label{mamma}
\eeq

The Green's functions entering the expansion (\ref{exp}) have to be computed
in the `unbroken phase' {\it i.e.} for vanishing background fields
$\Phi_i^c=\{H_{1,2}^0, {H_{1,2}^0}^*, a_\mu\}$. Taking $\Phi_i^c=0$ in eq.
(\ref{mamma}), one gets
\beq
\left.\partial_\mu \frac{\delta {\cal O}}{\delta a_\mu}\right|_{\Phi_i^c=0} =
0,
\label{cc}
\eeq
while, differentiating with respect to $H_i^0$,
\beq
-i\frac{e}{2} \left. \frac{\delta {\cal O}}{\delta
H_1^0(y)}\right|_{\Phi_i^c=0}
\delta^{(4)}(x-y)
+ \left.\partial_\mu \frac{\delta^2 {\cal O}}{\delta a_\mu(x)
\delta H_1^0(y)}\right|_{\Phi_i^c=0} = 0
\label{dd}
\eeq
and similar  identities are obtained differentiating with respect to the other
fields.

Now, let's indicate with $\overline{\Phi}_i^c =
\{\overline{H}_{1,2}^0, {\overline{H}_{1,2}^0}^*, \overline{a}_\mu=0\}$
the `bubble' solution of
the field equations discussed above, and with
${\overline{\Phi}_i^c}^\prime =
\{{\overline{H}_{1,2}^0}^\prime, {{\overline{H}_{1,2}^0}^\prime}^*,
\overline{a}_\mu^\prime = -1/e \: \partial_\mu \theta\}$
 the solution obtained
from the former by a gauge transformation.
Then, we consider the functional ${\cal O}$ on the background
${\overline{\Phi}_i^c}^\prime$  and expand it according to eq. (\ref{exp}).
Since $\overline{a}_\mu^\prime \neq 0$ in this case also Green
 functions involving gauge
fields in the external legs will contribute to the expansion whereas, if we
consider the background $\overline{\Phi_i}^c$, only the Green
functions with neutral
Higgses in the external legs contribute. However, by means of the above
identities, one can easily show that there are cancellations among these new
contributions, so that the result is the same as for   $\overline{\Phi_i}^c$.
Indeed, at the
linear order in $\overline{a}_\mu^\prime$ in the expansion we have
\beq
\int d^4x \left.\frac{\delta {\cal O}}{\delta a_\mu (x)} \right|_{\Phi_i^c=0}
\overline{a}_\mu^\prime(x) = \frac{1}{e} \int d^4x \partial_\mu \left.
\frac{\delta {\cal O}}{\delta a_\mu (x)} \right|_{\Phi_i^c=0} \theta(x) = 0
\eeq
where the last equality is due to eq. (\ref{cc}).
At the linear order in ${\overline{H}_1^0}^\prime$ we have
\beq
\int d^4x  \left. \frac{\delta {\cal O}}{\delta H_1^0(x)}\right|_{\Phi_i^c=0}
 {\overline{H}_1^0}^\prime(x) = \int d^4x  \left.
\frac{\delta {\cal O}}{\delta H_1^0(x)}\right|_{\Phi_i^c=0}
 \overline{H}_1^0(x)(1-i\theta(x)/2) + O(\theta^2).
\label{pappa}
\eeq
The contribution $O(\theta(x))$ is canceled by the ${\overline{H}_1^0}^\prime
\overline{a}_\mu^\prime$ term in the expansion by virtue of eq. (\ref{dd}),
\beq
\begin{array}{l}
 \ds \int d^4x d^4y \left.\frac{\delta^2 {\cal O}}{\delta a_\mu(x)
\delta H_1^0(y)}\right|_{\Phi_i^c=0} {\overline{H}_1^0}^\prime(y)
\overline{a}_\mu^\prime(x) =\\
\\ \ds
\frac{1}{e}\int d^4x d^4y
\left.\partial_\mu \frac{\delta^2 {\cal O}}{\delta a_\mu(x)
\delta H_1^0(y)}\right|_{\Phi_i^c=0} \overline{H}_1^0(y) \theta(y) +
O\left(\theta^2\right)=\\
\\ \ds
\frac{i}{2} \int d^4x \left.\frac{\delta {\cal O}}{\delta H_1^0(x)}
\right|_{\Phi_i^c=0} \overline{H}_1^0(x) \theta(x) +O\left(\theta^2\right).
\end{array}
\eeq
The contribution of n-th order in $\theta$ in eq. (\ref{pappa}), is canceled
by the terms coming from the  ${\overline{H}_1^0}^\prime
\left(\overline{a}_\mu^\prime\right)^i$ ($i=0\ldots n$) orders in the
expansion, due to the
appropriate Ward identities obtained by differentiating eq. (\ref{dd}) with
respect to $a_\mu$.

This result generalizes to the other terms of the
expansion: all the contributions of the same order in the primed Higgs fields
and of any order in $a_\mu^\prime$ in the expansion for
${\cal O}[{\overline{\Phi}_i^c}^\prime]$ sum up to the corresponding term of
 the same
order in the non-primed Higgs fields in the expansion for
${\cal O}[\overline{\Phi}_i^c]$, so that
\beq
{\cal O}[{\overline{\Phi_i^c}}^\prime] = {\cal
O}[\overline{\Phi}_i^c].
\eeq

Since in practical applications one has to truncate the expansion to a
finite order in the fields, it  is clear from the above discussion that this
can be done in a sensible way only if we consider the background
$\overline{\Phi}_i^c$, where the classical gauge field is zero, {\it i.e.} in
the unitary gauge.
\vskip 1cm
{\large\bf 3. Computation of the currents.}
\vskip 0.5 cm

We will now apply the method illustrated in the previous paragraph to the
computation of $\langle J^\mu_i (z) \rangle_{\Phi^0_i(x)}$, the expectation
value of the current operator
$J^\mu_i(x)$ ($i = H_{1,2}^0,\:H^{\pm},\: t_{L,R},\:b_{L,R}$)
on the background
$\bar{\Phi}^c_{j, k}=\{H_1^0,\: {H_1^0}^*,\:H_2^0,\:{H_2^0}^*\}$, which, for
definitness, we assume to be the `bubble' solution to the field equation of
motion discussed in the previous paragraph.

In the following, we will assume that there are no chemical
potentials. Then, it is easy to realize that a necessary condition for the
presence of non-vanishing expectation values for the currents is a non trivial
space-time dependence of the background. Indeed, if $H_{1,2}^*(x)$ were
constant then the only  two vectors available for the construction of
$\langle J^\mu_i (z) \rangle_{\Phi^c_j}$ would be $z^\mu$ and $u^\mu$ (the
quadrivector which identifies the motion of the thermal reference). By
using translational invariance, we are left with
$\langle J^\mu_i (z) \rangle_{\Phi^c_j} \sim {\rm const} \times u^\mu$, but
this is forbidden by CPT.

Also, by applying a charge congiugation transformation both to
the operator and to the background, we see that
\beq
\langle J^\mu_i (z) \rangle_{\Phi^c_j(x)} = -
\langle J^\mu_i (z) \rangle_{{\Phi^c_j(x)}^*},
\eeq
{\it i.e.} the expectation value for the current would vanish also if the
background were real.
{}From these two considerations one can conclude that the expectation values we
are looking for, will depend on {\it space-time derivatives} of a {\it complex
background}, so that the two phases $\theta_1(x)$ and $\theta_2(x)$ discussed
previously, (or $\delta(x)$ and $\chi(x)$), constrained by (\ref{aa}),
will play a crucial role.

Computing the expectation value as a functional expansion on the background,
as in eq. (\ref{exp}), the problem is reduced to the calculation of 1PI
Green's functions with one $J^\mu_i(x)$ insertion in the {\it unbroken} phase,
as in eq. (\ref{1p1}).

We will work in the imaginary time formalism, so that our expressions for the
Green's functions and for the currents are valid in the reference frame of the
thermal bath, where the bubble wall is moving.

The zero order of the expansion in eq. (5)
vanishes since it is given by the 1PI part of
$\langle J^\mu_i (z) \rangle_{\Phi^c_j=0}$ which is zero since translational
invariance holds in the unbroken phase.

The linear order is also zero, as can be realized observing that
in the unbroken phase the theory has still a {\it global}
$U(1)$ invariance, whereas
$\langle J^\mu_i (z) \hat{H}_k^0(y) \rangle_{\Phi^c_j=0}$ is not a singlet
under such symmetry.

The first non-vanishing  contributions then come from the quadratic terms
{\it i.e.}
\beq
\langle J^{\mu}_i(z) \rangle = \frac{1}{2}\sum_{j,k} \int d^4x\: d^4y
\left.\frac{\delta^{3}
\Gamma\left[\bar{\Phi}^c_j,\:\Delta_\mu^i\right]}{\delta \bar{\Phi}^c_j(x)
\delta \bar{\Phi}^c_k(y) \delta \Delta_{\mu}^i(z)}\right|_{\bar{\Phi}^c_{j,k}=
\Delta_{\mu}=0} \bar{\Phi}^c_j(x) \bar{\Phi}^c_k(y).
\label{quad}
\eeq

Using again the global $U(1)$, one can see that the only
possible non-vanishing  contributions to  the above expression come
from the 1PI part of
 $\langle J^\mu_i (z) \hat{H}_1^0(x) \hat{H}_2^0(y) \rangle_{\Phi^c_j=0}$, or
$\langle J^\mu_i (z) \hat{H}_i^0(x) {\hat{H}_i^0(y)}^* \rangle_{\Phi^c_j=0}$
($i=1,\:2$). These Green's functions will now be computed as usual in
perturbation theory, and the result inserted into the expression (\ref{quad}).

We start with the Higgs currents
\mbox{$J_{H_i}^{\mu} = i \left(H_i^{\dag} {\cal D}^{\mu} H_i -
{\cal D}^{\mu} H_i^{\dag} H_i\right)$}.
In the case of the neutral Higgses a contribution  already appears
at the tree level, see Fig.1. Inserting this contribution into the
expression (\ref{quad}) we get the classical current
\beq
\langle J_{H_i^0}^{\mu}(z) \rangle^{(1)} = - 2 \:
v_i^2(z)  \: \partial^{\mu} \theta_i
(z).
\label{clas}
\eeq
Note that, as expected, the individual Higgs currents are zero if the phases
are
constant.
Moreover, from the equation of motion
(\ref{aa}), we see that
the total Higgs hypecharge current vanishes at the tree level.

\begin{figure}
\leavevmode
\epsfxsize=5.cm
\epsfysize=5.cm
\centerline{\epsfbox{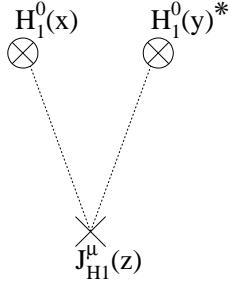}}
\caption{The tree level contribution to the neutral Higgs current.}
\end{figure}

At one loop there is a contribution to the neutral and charged Higgs currents
given by graphs like that in Fig. 2.

\begin{figure}
\leavevmode
\epsfxsize=5.cm
\epsfysize=5.cm
\centerline{\epsfbox{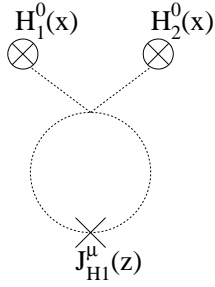}}
\caption{The 1-loop contribution to the neutral and charged Higgs currents.}
\end{figure}

Since the computation has to be performed in the unbroken phase, we must use
resummed propagators for the Higgs fields in order to deal with the IR
divergencies \cite{IR}. In the unbroken phase, the Higgs spectrum contains two
complex electrically neutral fields and two charged ones. At the tree level,
the squared masses of one of the neutral states and of one of the charged ones
are negative, since the origin of field space becomes a minimum of the
effective potential only after the inclusion of the finite temperature
radiative
corrections.
The resummation can be achieved by  considering the propagators for the
eigenstates of the thermal mass matrix, which has positive eigenvalues given by
\beq
M^2_{1,2}(T) = \frac{m_1^2(T) + m_2^2(T) \mp
\sqrt{\left(m_1^2(T)-m_2^2(T)\right)^2
+ 4 \:m_3^4(T)}}{2},
\eeq
where the $m_i^2(T)$ are the thermally corrected mass parameters of the
potential (\ref{pot}), $m_1^2(T)\simeq m_1^2 + 3 g^2 T^2 /16$,
 $m_2^2(T)\simeq m_2^2 +  h_t^2 T^2 /4$, while $m_3^2(T)$ receives only
logarithmic corrections in $T$, which were computed in ref. \cite{scpv}.
Correspondingly, the neutral complex eigenstates are given by
\beq
\left\{
\begin{array}{ccccc}
h &= &\cos\alpha \: H_1^0 &+& \sin \alpha \:{H_2^0}^*\\
&&&&\\
H &= &-\sin\alpha \:H_1^0& + &\cos \alpha \:{H_2^0}^*\\
\end{array},
\right.
\eeq
where
\beq
\sin 2\alpha = \frac{2 m_3^2(T)}{M_1^2(T) - M_2^2(T)}.
\label{alfa}
\eeq
Completely analogous formulae hold for the charged eigenstates.

The loop in Fig. 2 gives
\beq
\frac{\delta^3 \Gamma}{\delta \Delta^1_\mu(z) \delta H_1^0(x) \delta H_2^0(y)}
=
- i \:\lambda_5 \:\sin 2\alpha \:\:\delta^{(4)}(x-y) \int \frac{d^4q}{(2
\pi)^4} e^{i q\cdot(z-x)} {\rm H}^\mu(q) ,
\eeq
where the function $ {\rm H}^\mu(q)$ is the analytical continuation to
Minkowski external momentum $q$ of the finite temperature integral,
\beq
 {\rm H}^\mu(\tilde{q}) = \is dp \left\{ \frac{
p^\mu}{\left[p^2+M_1^2(T)\right] \:\left[(\tilde{q}-p)^2 + M_2^2(T)\right]}
 - \left((M_1^2(T) \leftrightarrow
M_2^2(T)\right)\right\},
\label{UV}
\eeq
where $\tilde{q}_4 = i q_0$, $\tilde{q}_i = q_i$,
$p_0 = 2\pi n_p T$, $p^2=p_0^2 + |\vec{p}|^2$, and
\[\is dp \equiv \sum_{n_p} \int \frac{d^3\vec{p}}{(2\pi)^3}.\]
Note that the only quartic coupling contributing to the Higgs currents is
$\lambda_5$, whereas there are no contributions from $\lambda_6$ and
$\lambda_7$. Moreover, the contribution vanishes if there is no mixing
$m_3^2(T)$ between the two Higgs gauge eigenstates (see eq. (\ref{alfa})).

It is easy to check that Green function in (\ref{UV}) is ultraviolet finite,
due to the minus sign between the two terms in the integrand.
After some algebra, the integral in $p$ can be casted into the form
\beq
\begin{array}{cr}
&\ds {\rm H}^\mu(\tilde{q}) =
\left.\frac{i}{2} \tilde{q}^\mu \right[(\tilde{q}^2+M_1^2(T)) \:\:\is \:dp
\ds \frac{1}{(p^2+M_2^2(T))((\tilde{q}-p)^2 +M_1^2(T))
((\tilde{q}+p)^2 +M_1^2(T))}\\
&\\
& \ds - \left. \left((M_1^2(T) \leftrightarrow
M_2^2(T)\right)\right].
\end{array}
\eeq
Now, the scale of the external momentum $q$ is set by the bubble wall width,
which, in the case of interest for us of thick bubble walls, is given by
$L_w^{-1}\simeq T/100- T/10$ \cite{thick}. Then, since we have
$\tilde{q}^2\ll M_{1,2}^2(T)\ll (2 \pi T)^2$, we approximate the integral by
neglecting the $\tilde{q}$ dependence in the integrand and by considering only
the
zero Matsubara mode for $p_0$. Inserting the result into the quadratic term of
the expansion (\ref{exp}) we obtain
\beq
\langle J_{H_1^0}^{\mu}(z) \rangle^{(2)} =
\langle J_{H_2^0}^{\mu}(z) \rangle^{(2)} \simeq \frac{\lambda_5}{8 \pi}
\frac{T\: m_3^2(T)}{(M_1(T)+M_2(T))^3}
\:\partial^\mu \left[v_1(z)v_2(z) \sin \delta(z)\right],
\label{h1l}
\eeq
 Each charged Higgs gets a
contribution equal to that of the neutral Higgs belonging to the same doublet.
As expected, also the Higgs currents vanish in the limit of vanishing
$v_i(z)$.

The first contribution to the fermionic currents arises at one loop and is
given by graphs like that in Fig. 3.
Once inserted into the expansion
(\ref{exp}), they give, in the case of the left handed top,
\beq\label{ht}
\langle J^{\mu}_{t_L}(z)\rangle^{(3)}=
i h_t^2\int d^4x\:
d^4y \:{\rm Im}\left(H_2^0(x){H_2^0(y)}^*\right)\:
{\cal G}^\mu(x,y,z),
\eeq
where $h_t$ is the top Yukawa coupling and
\beq
{\cal G}^\mu(x,y,z) =\int\frac{d^4l}{(2\pi)^4} \frac{d^4m}{(2\pi)^4} e^{i
\:l(x-z)} e^{i\:m(y-z)} {\rm G}^\mu(m,l)
\label{gg}
\eeq
is the
Green function corresponding to the diagram in Fig. 3, with
\beq
{\rm G}^\mu(\tilde{m},\tilde{l})= \:\:\is\:dk
\frac{{\rm Tr}\left[(\not \! k+\not \! \tilde{m})
\not \! k (\not \! k - \not \! \tilde{l})\gamma^\mu\right]}{(k+\tilde{m})^2 k^2
(k-\tilde{l})^2},
\label{G}
\eeq
and the zero component of the fermionic loop momentum is $k^0 = (2 n +1)\pi T$.
\begin{figure}
\leavevmode
\epsfxsize=5.cm
\epsfysize=5.cm
\centerline{\epsfbox{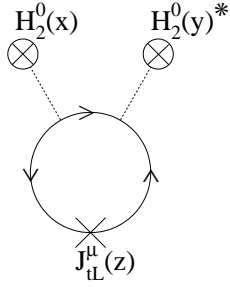}}
\caption{The 1-loop contribution to the left-handed top current.}
\end{figure}

Contrary to the case of the scalar loop, this fermionic loop integral is now
infrared finite, since the zero component of the loop momentum never vanishes.
As a consequence, we
will work with the tree level fermion propagators, without of
resummation.

Again, in presence of a thick wall background we can perform a high temperature
expansion. This is achieved by approximating ${\rm
G}^\mu(\tilde{m},\tilde{l})\simeq
{\rm G}^\mu(\tilde{m},0) +{\rm G}^\mu(0,\tilde{l})$ (note that
${\rm G}^\mu(\tilde{m},\tilde{l})= -
{\rm G}^\mu(\tilde{l},\tilde{m})$), and neglecting terms of order
$\tilde{l}^2/T^2$ in the
computation of ${\rm G}^\mu(0,\tilde{l})$.

Now the integral has a logarithmic ultraviolet divergence, so renormalization
is needed (see also the next section). In the $\overline{MS}$ scheme we obtain
\beq
\langle J^{\mu}_{t_L}(z)\rangle^{(3)}\simeq -\frac{h_t^2}{\pi^2}\:
v_2^2(z)\partial^{\mu}{\theta}_2(z)\:\left(\log\left(\frac{\mu^2}{A_f\:T^2}
\right) +\frac{3}{2}\right),
\label{jl}
\eeq
where $A_f=\pi^2 \exp(3/2 \:-2\gamma_E)\simeq 13.944$.

Eq. (\ref{jl}) shows the expected dependence on $h_t^2$ and $v_2^2(z)$
which, in comparison to the original result given in ref. \cite{CKN},
gives a suppression factor ${\cal O}(h_t v_2/\pi\:T)^2$.

A graph similar to that in Fig. 1 for the right handed top quark leads to a
contribution to $\langle J^{\mu}_{t_R}(z)\rangle$ given by $\langle
J^{\mu}_{t_R}(z)\rangle^{(1)} = - \langle J^{\mu}_{t_L}(z)\rangle^{(1)}$.
For the
other fermion species, one finds analogous results, in which $h_t$ is replaced
by the appropriate Yukawa coupling, and $v_2(z)$ ($v_1(z)$) appears for the up
(down)-type fermions.

A contribution to $\langle J^{\mu}_{t_L}(z)\rangle$ proportional to
${\rm Im} (H_1^0 H_2^0) = v_1 v_2 \sin\delta$, and to $h_t^2$,
appears at two loops, given by the graph in Fig. 4.
The corresponding integral is
\beq
I(k) = \is dq \frac{1}{( (q+k)^2 + M_1^2(T) )( q^2+M_2^2(T)) } {\rm
G}^\mu(-q-k,q)
\eeq
where $q$ is the bosonic loop variable and ${\rm G}^\mu(m,l)$ has been defined
in (\ref{G}).
Computing the fermionic integral in the approximation described  above,
we are left with a ultraviolet finite bosonic integration which  receives
important contributions only from the zero Matsubara mode. Inserting the
result into the expansion we obtain, again in the high temperature
approximation\footnote{Note a factor 6 of difference with respect to the
result of ref. \cite{expansion}},
\beq
\langle J_{t_L}^{\mu}(z)\rangle^{(4)}\simeq -\frac{\lambda_5 \: h_t^2 }
{64 \pi^3}
\frac{T\: m_3^2(T)}{(M_1(T)+M_2(T))^3}\left(\log\left(\frac{\mu^2}{A_f\:T^2}
\right) +\frac{3}{2}\right)
\partial^\mu \left[v_1(z)v_2(z) \sin \delta(z)\right].
\label{fatticunt}
\eeq
\begin{figure}
\leavevmode
\epsfxsize=5.cm
\epsfysize=5.cm
\centerline{\epsfbox{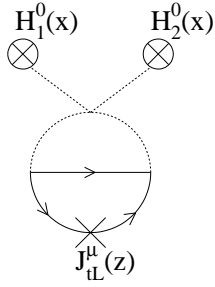}}
\caption{The 2-loop contribution to the left-handed top current. The scalar
internal lines can be either neutral or charged Higgs fields.}
\end{figure}

As for the one-loop result, the two loop contribution with neutral Higgses
in the internal lines for
$\langle J^{\mu}_{t_R}(z)\rangle$ is opposite to the one for
$\langle J^{\mu}_{t_L}(z)\rangle$.
This is no longer true when the graphs with charged Higgses in the internal
lines are taken
into account. In such a case, $\langle J^{\mu}_{t_R}(z)\rangle$ gets a
a contribution equal to eq. (\ref{fatticunt}), whereas
the result for $\langle J^{\mu}_{t_L}(z)\rangle$ is analogous but proportional
to $h_b^2$ instead of $h_t^2$.
The charged Higgs loops give also rise to a non-vanishing left handed bottom
density opposite to eq. (\ref{fatticunt}).

When applying our calculations
to the interesting case of spontaneous baryogenesis, one has to remember
that the typical values of $v_i(x)/T$ are smaller than one, as required
in regions of the bubble wall where sphalerons are still active. This
fact enables us to stop the expansion series at the second order, being
the remaining terms suppressed by powers of $v_i(x)/T$ with respect to
it.

\newpage
{\large\bf 4. Current renormalization.}
\vskip 0.5 cm
In the previous section we have seen that in order to calculate the
finite temperature average of the particle currents,
the expansion (\ref{exp}) can be used,
whose coefficients
are the 1PI Green's functions with the insertion of current operators
$J^\mu_i$, computed in the
unbroken phase. With $J^\mu_i$'s we have indicated the current for the i-th
particle, expressed in terms of the renormalized fields.

In general, the
computation of Green's functions with the insertion of a composite operator
requires the introduction of new counterterms besides those necessary for the
renormalization of the basic lagrangian (see ref. \cite{collins}  for a
thorough discussion). For example, the one loop computation of the Green
function ${\cal
G}^\mu(x,y,z)=\langle J_{t_L}^\mu(z) H_2^0(x) H_2^{0,\star}(y)\rangle$,
defined in eq. (\ref{gg}), gives a divergent part
(in the $\overline{MS}$ renormalization scheme, and in the momentum
space)
\begin{equation}
\tilde{\rm G}^{\mu,div}(m,l)=
+\frac{1}{8\:\pi^2}h_t^2 \frac{1}{
\Gamma(3)}\frac{1}{\varepsilon}\left(m^\mu-l^\mu\right).
\end{equation}
where $m$, $l$ are the momenta associated to the external Higgs fields
and $\varepsilon = 4-n$.
This divergence may be expressed as
\beq
 {\cal G}^{\mu,div}(x, y, z)=-\frac{1}{16\:\pi^2}\frac{1}{\varepsilon}
\:\langle J^\mu_{H_2^0}(z) H^0_2(x) H^{0 *}_2 (y)\rangle
\eeq
{it i.e} it is proportional to the {\it tree level} Green function involving
the $H_2^0$ Higgs field current, corresponding to Fig. 1.

This  is an example of a quite general phenomenon: a
renormalized current $\left[J^\mu_i\right]$,  can be defined as a linear
combination of the currents $J^\mu_j$
\beq
\left[J^\mu_i\right] = \sum_j (\delta_{ij} + \delta Z_{ij}) J^\mu_j
\label{mixing}
\eeq
in such a way that all the possible Green's functions with
 $\left[J^\mu_i\right]$ insertions are finite.
In the example discussed above, we have $\delta Z_{L2}
= -\mu^2 h_t^2/(16 \pi^2\:\varepsilon)$. The $\delta Z_{ij}$ are
in general new renormalization constants which are not contained in the
counterterm lagrangian, so that new renormalization conditions are needed.

An important exception to this is the case in which $J^\mu$ is a conserved
current for the renormalized lagrangian
\beq
{\cal L}={\cal L}_{basic}+{\cal L}_{ct},
\label{split}
\eeq
where ${\cal L}_{basic}$ is expressed in terms of renormalized quantities and
${\cal L}_{ct}$ is the counterterm Lagrangian. Let's consider here the simple
case of a lagrangian containing only left and right handed top quarks and  the
neutral Higgses, which will be enough for discussing the renormalization of
the Green's functions relevant for this paper.

The conserved current under the group $U(1)$  is then
\beq
J^{\mu}+\delta J^{\mu}= q_L \left(1+\delta Z_L\right)\:J^{\mu}_{L}+
q_R \left(1+\delta Z_R\right)\:J^{\mu}_{R}+
q_1 \left(1+\delta Z_1\right)\:J^{\mu}_{H^0_1}+
q_2 \left(1+\delta Z_2\right)\:J^{\mu}_{H^0_2},
\label{cur}
\eeq
with $q_2=q_R-q_L=-q_1=1/2$ and
$\delta J^{\mu}$ is the part  of the current coming from ${\cal L}_{c.t.}$.
The $\delta Z_{\alpha}$ ($\alpha=L,\:R,\:1,\:2$) are now the
wave-function renormalization constants for the different fields, coming from
${\cal L}_{ct}$.
In this case, being the current conserved, no new counterterms besides those
already present in (\ref{cur}) (and hence in ${\cal L}_{ct}$)
 are needed \cite{collins}, {\it i.e.}
\beq
J^\mu + \delta J^\mu = \left[J^\mu\right]
\equiv q_L \left[J^\mu_L\right] + q_R \left[J^\mu_R\right] + q_1
\left[J^\mu_{H_1^0}\right] +  q_2\left[J^\mu_{H_2^0}\right].
\label{theo}
\eeq

Recalling this property of the conserved current it is possible to reduce the
number of linearly independent renormalization constants $\delta Z_{ij}$
(where now $i= L,\: R,\:1,\:2$)
necessary in order to define the renormalized currents according to eq.
(\ref{mixing}). Indeed, from (\ref{theo}), recalling
(\ref{mixing}) and (\ref{cur}), the number of new linearly independent
renormalization constants
$\delta Z_{ij}$'s is reduced from 16 to 8.

The most general renormalization  matrix is then given by
\beq
\left(
\begin{array}{cccc}
1 +\delta Z_{LL}& \delta Z_{R}- \delta Z_{RR}  &  \delta Z_{L1} &
\delta Z_{L2}\\
\delta Z_{L}- \delta Z_{LL} & 1+  \delta Z_{RR} & - \delta Z_{L1} &
-\delta Z_{L2}\\
 \delta Z_{1L} &  \delta Z_{1R} & 1 + \delta Z_{11} &  \delta Z_{12}\\
\delta Z_{1L} + \delta Z_{LL} - \delta Z_L &
\delta Z_{1R} -\delta Z_{RR} +\delta Z_R  &
\delta Z_1 +\delta Z_{11} +\delta Z_{L1} & 1 +\delta Z_2+\delta Z_{L2} +
\delta Z_{12}
\end{array}
\right)\:
\label{matrix}
\eeq
We see then that, in general, eight extra renormalization conditions are
required in order to define the set of renormalized currents
$\{\left[J^\mu_i\right]\}$.

The only divergent graph encountered in sect. {\bf 3}, is the
fermionic one loop (see eq. (37)), so, for the present application,
and to our order of computation, it is enough to compute the matrix
(\ref{matrix}) only to order $(g)^0 h_t^2$.
To this order, only one of the eight extra renormalization constants is non
zero, $\delta Z_{L2}$ and moreover we obtain, again from (\ref{theo}), that
 $\delta Z_{L2}= -\delta Z_2$ (+ higher order terms), so that no new
renormalization condition is needed.

The mixing matrix is then given, at this order, by
\beq
\left(
\begin{array}{c}
\left[J^{\mu}_{L}\right]\\
\left[J^{\mu}_{R}\right]\\
\left[J^{\mu}_{1}\right]\\
\left[J^{\mu}_{2}\right]\end{array}\right)=
\left(
\begin{array}{cccc}
1 & \delta Z_{R} & 0 &-\delta Z_{2}\\
\delta Z_{L} & 1 & 0 &\delta Z_{2}\\
0 & 0 & 1 & 0\\
-\delta Z_{L} & \delta Z_{R} & 0 & 1
\end{array}
\right)
\left(
\begin{array}{c}
J^{\mu}_{t_L}\\
J^{\mu}_{t_R}\\
J^{\mu}_{1}\\
J^{\mu}_{2}
\end{array}
\right),
\label{final}
\eeq
where $\delta Z_{L2} = -\delta Z_2=(-\mu^{2}h_t^2/16\:\pi^2\:\varepsilon)$,
$\delta Z_L=\delta Z_R=(-\mu^{2}h_t^2/32\:\pi^2\:\varepsilon)$.

Working at higher order, we would need new renormalization conditions, defining
the physical currents $\left[J^\mu_i\right]$ at a certain
scale. Due to the interactions, at a different scale
the renormalized currents will be
given by different mixings of  the unrenormalized ones, so that, in general,
the definition of `pure' ({\it i.e.} non-mixed) currents  is scale dependent.
As we have already stressed, this is not the case for the conserved
currents, which do not suffer any mixing and do not require any new
renormalization condition.

As we will discuss in the following section, the only relevant currents in the
limit of local thermodynamical equilibrium are just the conserved ones, so
that in this case, the state of the  system is completely determined without
the need of introducing extra renormalization conditions.

\vskip 1cm

{\large\bf 5. The limit of local equilibrium.}
\vskip 0.5cm
Up to this point, we have computed the expectation values of the individual
particle numbers starting from a partition function of the form
\beq
Z[J_i, \Delta_j] = {\rm Tr}\: \exp\left[-\beta \left(H  +
 \int \Delta_j
N_j\right)\right]
\label{partition}
\eeq
where $H$ is the Hamiltonian of the system and we are now considering only the
zero components ($N_j \equiv J^0_j$) of the currents  considered in the
previous paragraphs. We assume that all the chemical potentials are zero.

If we take generic values for the sources $J_i$ and
$\Delta_j$ the partition function (\ref{partition}) will not correspond to the
one describing the limit of thermodynamical equilibrium, in which the
individual particle numbers assume values such that the free energy of the
system is minimized. In the case under consideration of a space-time dependent
Higgs field background,  the limit of  {\it local} equilibrium is achieved only
if all the
interaction rates are much higher that the inverse of the typical space-time
scale  of the variation of the background, {\it i.e.} the bubble wall width.
This condition is fulfilled in reality only by a few interactions (the gauge
flavour conserving ones, the top Yukawa interactions, and possibly some
Higgs-Higgs interactions), however it is anyway instructive to consider first
this limit before discussing the more realistic situation in which some of the
interactions are out of equilibrium inside the bubble wall.

The free energy of the system is obtained by performing the Legendre
transformation
of the functional defined in eq. (\ref{start}) with respect to the sources
$\Delta_j$ of the density operators,
\beq
\begin{array}{cl}
\ds {\rm F}\left[\Phi^c_i,\:{\cal N}_j^c\right]& = W\left[J_i,\:\Delta_j
\right]
 - \sum_i \int d^4x J_j(x) \Phi^c_j(x) - \sum_j \int d^4x \Delta_j(x) {\cal
N}_j^c(x)\\
&\\
&\ds
=\Gamma\left[\Phi_i^c, \Delta_j\right] - \sum_j \int d^4x \Delta_j(x) {\cal
N}_j^c(x)\\
&\\
&\ds =\langle H \rangle - T S.
\end{array}
\label{free}
\eeq
The last equality in the above equation is straightforwardly derived
recalling that the entropy $S$ is given by $S=-\partial W/\partial T$ and that
$W=-T \log Z$ (we are taking the three dimensional volume $\Omega = 1$) .

The classical densities ${\cal N}_j^c(x)$ are defined as usual,
\beq
{\cal N}_j^c(x) = \left.\frac{\delta \Gamma[\overline{\Phi}_i^c, \Delta_j]}
{\delta
\Delta_j(x)}\right|_{\Delta_j = 0},
\label{aaa}
\eeq
where the background is, as before (eq. (\ref{fe})), the solution of the field
equations with
appropriate boundary conditions, {\it i.e.} we have $J_i(x)=0$ in eq.
(\ref{partition}).

Now, the local equilibrium situation on the background
$\overline{\Phi}_i^c(x)$ is obtained once all the  processes have driven the
free energy to its minimum possible value. This condition corresponds to
a system of $M$
equations (where $M$ is the number of processes in equilibrium) of the form
\beq
\delta^{(l)}  {\rm F}\left[\Phi^c_i,\:{\cal N}_j^c\right] =
 \sum_j \nu_j^{(l)}\:
\frac{\delta  {\rm F}\left[\Phi^c_i,\:{\cal N}_j^c\right]}
{\delta {\cal N}_j^c(x)} = -\sum_{j=1}^n \nu_j^{(l)} \Delta_j(x)
= 0\:\:\:\:\:\:\:\:\:\:\:(l=1,\ldots, M),
\label{chem}
\eeq
where $\nu_j^{(l)}$ is the multiplicity of the $j$-th particle in the $l$-th
process ({\it i.e.} in the process $A + 2B \leftrightarrow C+D$ we have
$\nu_A=-\nu_C=-\nu_D=1$, and $\nu_B=2$).

The system (\ref{chem}) is of course strictly analogous to the usual system of
equations that one has to solve to determine the equilibrium chemical
potentials. In this case, it restricts the set  of the linearly independent
sources to $\{\Delta_a^\prime\}$ ($a=1,\ldots, N$),
corresponding to the set of the $N$ conserved charges $N_a$ of the system. This
means that the only composite operators that appear in the {\it equilibrium}
partition function
\beq
Z_{eq}[J_i, \Delta_a^\prime] =\left. {\rm Tr}\:
 \exp\left[-\beta \left(H +
 \int \Delta_a^\prime
N_a\right)\right]\right|_{\Delta_a^\prime=0}
\eeq
are those corresponding to the conserved charges,
\beq
{\cal N}_a^c(x) = \left.\frac{\delta \Gamma_{eq}[\overline{\Phi}_i^c,
\Delta_a^\prime]}
{\delta
\Delta_a^\prime(x)}\right|_{\Delta_a^\prime = 0},
\eeq
where $\Gamma_{eq}$ is obtained from $Z_{eq}$ in the standard way.

After solving the
system (\ref{chem}), the  sources for the individual particle numbers,
$\Delta_i$, are expressed in terms of
the new sources $\Delta_a^\prime$ as
\beq
\Delta_i = \sum_a  q_a^{(i)}\: \Delta_a^\prime,
\eeq
where $q_a^{(i)}$ is the $a$-charge value of the particle $i$. Then, the
{\it equilibrium} values for the individual particle numbers are given by
\beq
{\cal N}_i^{eq}(x) = \sum_a  \left.\frac{\delta
\Gamma_{eq}[\overline{\Phi}_i^c,
\Delta_a^\prime]}
{\delta
\Delta_a^\prime(x)}\right|_{\Delta_a^\prime = 0}\:
\left(q_a^{(i)}\right)^{-1} =
\sum_a {\cal N}_a^c(x)  \left(q_a^{(i)}\right)^{-1}.
\label{eq}
\eeq

Concerning renormalization we see that
since now the only Green's functions
that have to be computed  are those involving the
conserved charges, no new renormalization conditions for the composite
operators are necessary when one is
dealing with the local equilibrium case.

However in many interesting applications not all the processes have rates
fast enough such that a complete local equilibrium  situation can be attained.
In this case one can consider an adiabatic approximation, in which the `slow'
processes are freezed out inside the bubble wall, while the fast ones are in
equilibrium. As a consequence, on the bubble wall only the equations
corresponding to such `fast' processess will contribute to the system
(\ref{chem}) and one can repeat the above analysis with
new effectively conserved charges now playing the role of the
$N_a$'s.

There is in this case a, mainly  conceptual, problem, since these
charges are now not conserved by the full theory, but only by the effective
one obtained by sending to zero all the couplings contributing to the `slow'
processes. Then they do in general mix one another under renormalization, as
was discussed in the previous section, so that the interpretation in terms of
physical currents will be scale dependent.

However, since the mixing
is due only to the small couplings which are neglected in the effective
theory, it will be in most of the practical applications a negligible effect.

\vskip 1cm
{\large \bf 6. Conclusions}
\vskip 0.5cm

Motivated by the recent interest on the possibility of generating the baryon
asymmetry in the early Universe during the electroweak phase transition
and, more in particular, by the so called spontaneous baryogenesis
mechanism in which the baryon asymmetry is  created inside
thick $CP$-violating and expanding bubble walls of the Higgs fields, we
have developed a general method to calculate the expectation value
of a composite operator in the presence of a generic nonvanishing
background for
 some of the fields of the theory.

The method is based on a functional expansion of the expectation value in
powers of the background fields $\Phi_i^c(x)$ which are a solution of the
equation of motion with properly chosen boundary conditions.
At any order  in the functional expansion, the expectation value is
determined by a 1P1 Green's function in the unbroken phase, which can be
computed in perturbation theory.

Using Ward identities, we have shown
that the expectation value for a gauge invariant operator is independent on
the gauge in which the background has been computed.

The renormalization of the currents has been also discussed,
stressing the crucial role played by the conserved currents
in the case of local equilibrium. In this limit, all the individual particle
currents are determined in terms of the conserved ones,
whose computation do
not require the introduction of new renormalization constants besides the ones
already contained in the renormalized lagrangian.

In practice, the system is well far away from such an idealized situation,
with the possible exception of the processes mediated by the top Yukawa and
the gauge flavour diagonal couplings. In this case, a better description of
the evolution of particle numbers can be achieved by  employing a system
of kinetic
equations. For each particle number, the values expressed by (\ref{eq})  would
then represent the infinite time limit of the corresponding rate equation,
\beq
\dot{\cal N}_i \sim \Gamma_{ij} ({\cal N}_j - {\cal N}_j^{eq}).
\eeq
As we have already stressed in the introduction, the dependence of such
equilibrium values on the coupling constants and on the value of the Higgs
background should not be confused with the dependence of the rates
$\Gamma_{ij}$ on these quantities.

The computation of the currents is of  course only the first step towards a
reliable determination of the final value of the baryon asymmetry in this
scenario, the major source of uncertainty being represented by the rate of the
baryon number violating processes inside and in front of the bubble walls.

In any case, due to its generality, we believe that the method illustrated in
this paper can be easily applied to similar situations of interest in
Cosmology, in which some of the fields acquire a  space-time
dependent classical value as for instance in inflationary models
\cite{Dolgov}, or in
presence of domain walls, as considered in
in ref. \cite{abel} in the case of the Next-to-Minimal
Supersymmetric Standard Model.

The problem of the computation of the currents on a CP violating background
has been addressed recently also by Huet and Nelson \cite{huet} for the case
of a fermionic axial current. Although
they adopt  a
semiclassical approach quite different from  the one discussed in this paper,
their results are in
quantitative agreement with ours. In particular the same dependence
on $h_t$ and $v(x)$ is found.

\vskip 0.5cm
\centerline{\bf Acknowledgements}

It is a pleasure to thank W. Buchm\"uller for a careful reading of the
manuscript and for providing us with helpful comments.
Useful discussions with A. Dolgov, F. Feruglio and A. Nelson are gratefully
acknowledged.

\end{document}